\documentclass[12pt]{iopart}
\usepackage{graphicx}% Include figure files
\usepackage{iopams}
\usepackage{subcaption}
\usepackage{dcolumn}% Align table columns on decimal point
\usepackage{bm}% bold math
\usepackage{xcolor}
\usepackage{soul}
\renewcommand{\deg}{\ensuremath{^{\circ}}}

\usepackage{titlesec}
\usepackage{relsize}
\makeatletter
\DeclareRobustCommand{\text}{%
  \ifmmode\expandafter\text@\else\expandafter\mbox\fi}
\let\nfss@text\text
\def\text@#1{{\mathchoice
  {\textdef@\displaystyle\f@size{#1}}%
  {\textdef@\textstyle\f@size{#1}}%
  {\textdef@\textstyle\sf@size{#1}}%
  {\textdef@\textstyle \ssf@size{#1}}%
  \check@mathfonts
  }%
}
\def\textdef@#1#2#3{\hbox{{%
                    \everymath{#1}%
                    \let\f@size#2\selectfont
                    #3}}}
\makeatother
\begin{document}
\BB 

\title[Comparison of tethered and self-propelled models of fish locomotion]{Performance comparison of tethered and self-propelled models of fish locomotion using unsteady thin airfoil theory}

\author{Anshul Nayak$^1$, Emad Masroor$^2$, and Hodjat Pendar$^3$}

\address{$^1$ Mechanical Engineering, Virginia Tech}
\address{$^2$ Department of Engineering, Swarthmore College}
\address{$^3$ Engineering Mechanics, Virginia Tech}
\ead{hpendar@vt.edu}

\begin{abstract}
Numerous experimental and computational studies have been conducted in the past few decades to understand the swimming performance of fish, and to identify the optimal kinematic strategies for a swimming fish or fish-like robot. Many of these studies model the swimmer in a `tethered' condition, in which the swimmer is held fixed while it is subjected to a free stream. Its performance is then quantified using power expenditure, thrust generation, and efficiency. However, the dynamics of a tethered swimmer are different from those of a self-propelled swimmer, whose performance is best measured using its steady-state swimming speed in still fluid and its efficiency. It is an open question whether the conclusions drawn from studies of tethered swimmers can be directly applied to free swimmers. In this study, we use an unsteady panel method to systematically compare the swimming performance of a tethered fin and that of a self-propelled fin attached to a virtual drag-producing body to investigate how their performance varies over a set of prescribed kinematics. After validating the numerical model against previous experimental results, we show how the pitch amplitude, heave amplitude and the phase offset between them affect the efficiency and thrust generation in the tethered case, and how they affect the speed and efficiency in the self-propelled case.  We find that the kinematic strategies that optimize the performance of a tethered swimmer do not necessarily optimize the performance of a self-propelled swimmer. 
\end{abstract}

\maketitle

%\tableofcontents
\section{Introduction}

% {Comments} 

The dynamics and propulsive performance of swimming fish are often investigated scientifically using tethered models of fish suspended in a flowing water channel \cite{anderson1998oscillating,VanBuren2019Scaling}. In experimental studies in particular, these `tethered' models allow the fish to be observed in a fixed frame of reference while water flows past it at a constant speed $U$, simulating the motion of a fish at a speed $U$ through still water. Even in computational studies of the fluid dynamics of swimmers, it is common to use a frame of reference attached to the swimming body while the surrounding fluid is imparted a constant speed $U$ \cite{michelin2009resonance,alben2009swimming,maerten2017soptimal}. 
These studies are based on the principle that the swimming dynamics of a tethered object in a free stream with speed $U$, using a Galilean transformation, are the same as the dynamics of the same object moving at a constant speed $U$ through a quiescent fluid. However, in studies with a tethered fin, the speed $U$ is an independent variable that can be specified, whereas a self-propelled swimmer attains a steady speed $U$ as a result of the balance between thrust and drag forces on it; in the latter case, the value of $U$ emerges from a particular experiment and cannot straightforwardly be specified. 

When a tethered tail fin --- without an attached body --- generates an average thrust $T$ for a given set of kinematic parameters (such as heave and pitch) when subjected to a free-stream velocity $U$, it is possible to design a drag-inducing upstream body shape such that, when the same tail fin is attached to this upstream body and allowed to swim freely in the streamwise direction in still water, it happens to reach an average velocity of exactly $U$. To accomplish this `matching' procedure correctly, one must choose an upstream body shape that experiences a drag force equal to $T$ when tethered in a free stream of velocity $U$. If, subsequently, a different set of kinematic parameters is to be studied, the average thrust in the tethered case will change, which will also change the average velocity attained by the self-propelled swimmer, rendering the comparison between the two cases invalid unless a different upstream body shape is selected. The new upstream body shape would have to be selected such that, when the fin is tethered to it and allowed to swim freely, it reaches the same velocity $U$ despite the modified kinematics being employed. This simple argument suggests that comparisons between tethered and self-propelled swimmers must be made with caution.

We therefore propose to answer the following question: Can the propulsive performance of a tethered swimmer subjected to streaming flow predict the self-propelled performance of the same swimmer in quiescent fluid? To what extent are the conclusions drawn from studies of tethered swimmers valid for self-propelled swimmers? The applicability of tethered-fin experiments to the design of propulsors has recently been called into question \cite{young2020cyber}, and a recent study has explored this problem by considering small variations in the flow speed past a tethered swimmer \cite{van_buren_flow_2018}. We are not aware of any systematic studies that explicitly compare the performance of tethered and self-propelled swimmers. In this work, we use the term `self-propelled swimmer' to refer to a body that is free to move in the streamwise direction under the influence of the thrust and drag forces it experiences; the body is artificially constrained to have no net motion in the transverse and rotational directions.

Here we focus on one particular problem that is often of interest to the scientific community and proceed to investigate it via both models: for a simultaneously heaving and pitching tail fin, what is the optimal phase lag between pitch and heave, $\phi$, that maximizes the swimmer's performance? This problem is particularly relevant to the question of tethered versus self-propelled swimmers because a consensus has developed that the optimal performance of simultaneously heaving and pitching air- and hydrofoils occurs at $\phi \approx 270 \deg$ \cite{isogai1999effects,read2003forces,boudreau_free-pitching_2019}. However, this conclusion has arisen out of measurements of the thrust coefficient $C_T$ and of the Froude efficiency $\eta$ of tethered airfoils, and it is not clear whether the kinematics that maximize $C_T$ or $\eta$ for a tethered swimmer would maximize the performance of a self-propelled one as well. Because a self-propelled swimmer at a constant speed has no net force on it, it is difficult to separate the thrust and drag forces acting on it \cite{schultz_power_2002}.

In this work, we explore various measures of the performance of both tethered and self-propelled swimmers using the same caudal fin model, and explore how each of these performance metrics vary with the kinematics imparted to the fin. In many species of fish, the majority of the thrust comes from the caudal fin (up to 90 percent \cite{bainbridge_caudal_1963,sfakiotakis_review_1999}). Therefore, we restrict our attention to caudal fin-driven swimming in which a fish uses periodic motion of its tail fin to propel itself forward, whereas its upstream body only contributes a drag force in the self-propelled case. We neglect the detailed fluid dynamics of the rest of the fish's body, and model only its tail fin using a two-dimensional flat plate of negligible thickness, actuated at its leading edge or peduncle. 

The rest of this paper is organized as follows. Section~\ref{sec:model} develops the computational model that we apply to both the tethered and the self-propelled case. In section~\ref{sec:performance}, we show how the performance of each model is quantified and evaluated, and in section~\ref{sec:validation}, we validate our numerical method against experimental work. Section~\ref{subsec: Tethered fin} presents the results for a tethered fin at a fixed background speed $U_0$, while section~\ref{subsec:freelyswimmingfin} presents the results for a self-propelled swimmer that develops a steady-state speed $\overline{v}$ in the presence of an assumed drag-producing body upstream of the caudal fin, which imparts a drag coefficient $\mu$ to its motion. Section~\ref{sec:conclusion} compares what we learn about the two cases and summarizes our conclusions regarding the applicability of the tethered-fin model to a self-propelled caudal fin-driven swimmer.

\section{Methods}
\subsection{Computational model}
\label{sec:model}
\begin{figure}[h]
    \centering
    \includegraphics[width=1\textwidth]{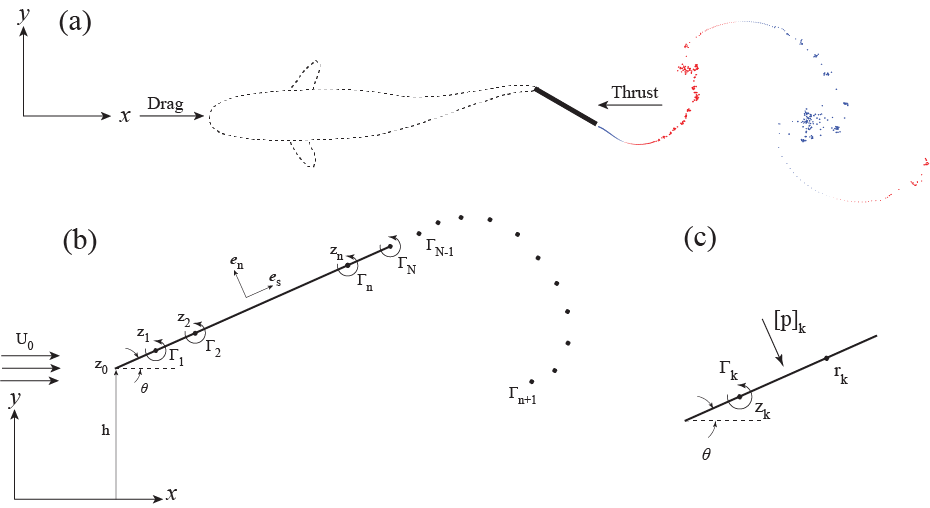}
    \caption{(a) A significant portion of the thrust is generated in the caudal fin to overcome  the drag force on the body. (b) The fin is divided into $n$ panels with a vortex element attached to the first quarter-chord of each panel. The strength of the vortex elements evolve with time. At each time step a vortex element is released from the trailing edge into the flow. (c) The no-penetration boundary condition is satisfied at $n$ collocation points $r_k$, which are at the three-quarter-chord of each panel.}
    \label{fig:panelmethodschematic}
\end{figure}

In this work, we model the caudal fin of a swimmer, shown schematically in fig.~\ref{fig:panelmethodschematic}(a), as a thin rigid plate with negligible mass immersed in an inviscid fluid, subject to an imposed heaving and pitching motion at its leading edge. The use of a rigid, rather than flexible, `fin' is a simplification of the true dynamics that significantly speeds up the numerical calculations by making the coupling between the solid body dynamics and the fluid dynamics one-way, instead of the two-way coupling that occurs when the fin is modeled as flexible. There is ample precedent in the literature for using rigid bodies as a first step in studying the mechanics of caudal fins; see, e.g., \cite{triantafyllou1993optimal,floryan2017scaling,van_buren_flow_2018}. 

Consider therefore a thin rigid plate of length $L$, actuated with a sinusoidal heave $h(t)$ and pitch $\theta(t)$ at its leading edge given by 
\begin{equation}
\label{eq:prescribedKinematics} 
h(t) = h_{o}  \sin (2\pi t), \qquad \text{and}  \qquad \theta(t) = \theta_{o}  \sin(2\pi t +\phi)
\end{equation}
respectively. In this model $\omega = 2\pi$ is the fixed non-dimensionalized frequency of oscillation and $\phi$ denotes a variable phase offset between pitch and heave. The plate is immersed in two-dimensional inviscid fluid with density $\rho_f$. Lengths and times are non-dimensionalized by $L$, $2\pi/\omega$ respectively.

Using thin airfoil theory, the fin is modeled by a series of bound vortex elements of time-varying strength placed at equal intervals on the fin \cite{eldredge2019mathematical}. Consistent with previous studies \cite{alben2009simulating,eldredge2019mathematical}, the wake of the fin is represented by a series of vortex elements that are released from the trailing edge. Following \cite{katz2001low}, the plate is divided into $n$ panels, and each panel is associated with an attached vortex element with an unknown and time-varying strength $\Gamma_j(t)$, $j=1,..,n$, at the quarter-chord of each panel (figure \ref{fig:panelmethodschematic}(b-c)). The free vortex sheet is represented by point vortices released at regular intervals $\Delta t$ from the trailing edge, and the strength of the point vortex released at each time step is determined according to Kelvin's circulation theorem. Once released into the fluid, each point vortex maintains its circulation in accordance with Helmholtz's Laws.

At time step $k$, the strengths of the $n$ attached vortices $\Gamma_j(t)$, $j=1,..,n$ and of the newly-shed vortex $\Gamma_{n+k}$ are unknown. To determine these unknowns, we satisfy the no-penetration condition at $n$ collocation points, located at the third quarter-chord point of each panel (fig.~\ref{fig:panelmethodschematic}(c)), as well as Kelvin's circulation theorem for the conservation of circulation inside a contour enclosing the plate and all of the shed vortices. Thus, we obtain a system of $n+1$ equations for the $n+1$ unknowns at each time step.

We use the same numerical method to model two different physical scenarios. In the first scenario, which we have called the `tethered' case, the flat plate is assumed to represent the caudal fin of a fish that is tethered to a point in space and then subjected to a free stream of speed $U_0$. In the second scenario, which we have called the `self-propelled' case, the flat plate is assumed to represent the caudal fin of a fish that is free to move in the streamwise direction under the influence of the horizontal forces generated by its swimming motion in otherwise stationary fluid. In this case, the speed developed by the swimmer is an output of the model, rather than an input. {Following the `virtual body' approach of \cite{akoz2021intermittent}, we assume that the anterior body is well streamlined, and its effect on the hydrodynamics of the fin is negligible; its only contribution is a drag force, proportional to the square of the horizontal speed of the body. Assuming the body moves only in the $x$-direction and has a nondimensionalized mass of $m$, the equation of motion in the horizontal direction is 
\begin{equation} 
 \label{eq:fin_equation2}
     m \ddot{x}  = T - D = T - \mu v^2,
\end{equation}
where  $T$ and $D$ are the thrust generated by the caudal fin and the drag force on the upstream body respectively. The mass, normalized by $\rho_f L^2$, is set equal to 2 throughout this study. The drag coefficient $\mu$ depends on the shape and size of the swimmer's body; in this work we have considered three different values for this coefficient: 0.025, 0.1 and 0.2 \cite{bilo1980simple, gazzola2014scaling, nesteruk2014shape} to cover a wide range of shapes.

The velocity of the fluid ${w}(z)$ at any location in the complex plane $z=x+iy$ is affected by both the bound and free vortex elements. The complex conjugate of the velocity $\overline{w}$ can be determined using the Biot-Savart formula \cite{jones2003separated}:
\begin{equation} 
\label{eq:fluid_vel} 
\overline{w}(z)=  U_{0}+\frac{1}{2\pi{i}}\sum_{j=1}^{N} \frac{\Gamma_{j}}{z - z_{j}},  \label{eq:vel_eqneq:fluid_vel} 
 \end{equation}
where $U_{0}$ is the velocity of the free stream, $\Gamma_{j}$ is the strength of the $j^{\text{th}}$ vortex element, and $ z_{j}$ is the location of the $j^{\text{th}}$ vortex element in the complex plane. For the self-propelled case, $U_0$ in equation (\ref{eq:fluid_vel}) can be replaced with the velocity of the body. To satisfy the no-penetration boundary condition, the velocity of the fluid and fin must be equal in the normal direction at the $j$ collocation points:     
\begin{equation}
{{w_{j}^{f}}}{} \cdot \hat e_{n} = {{w_{j}^{s}}}{} \cdot\hat e_{n} \qquad j = 1,..,n
\label{eq:vel_const}
\end{equation}
where $w_{j}^{f}$ and $w_{j}^{s}$ are the velocity of the fluid and fin at the $j^{th}$ collocation point. Combining
equation (\ref{eq:fluid_vel}) and equation (\ref{eq:vel_const})  and using the identity
$w \cdot \hat e_{n} =  Re(\overline{w} \hat e_{n})$ leads to:
\begin{equation}
Re\left[\hat e_{n}\left(U_{0}+\frac{1}{2\pi{i}}\sum_{j=1}^{N} \frac{\Gamma_{j}}{z_j^{s} - z_{j}}\right) \right] = {{w_{j}^{s}}}{}\cdot\hat e_{n}  \qquad j = 1,..,n 
\label{eq:no_penetration}
\end{equation}
where $z_j^{s}$ is the location of the $j^{th}$ collocation point. In equation (\ref{eq:no_penetration}), the strength of the bound vortices $\Gamma_1, \Gamma_2, ..., \Gamma_n$  and the strength of the nascent vortex released into the wake, $\Gamma_{N}$, are unknowns. The remaining $N-(n+1)$ vortices, whose circulations are $\Gamma_{n+1}, \Gamma_{n+2}, ..., \Gamma_{n+k-1}$, are free vortices whose circulations are known, having been fixed at the time of their shedding. Kelvin's circulation theorem indicates that the total circulation in the fluid, including the bound and free vortices, is conserved. Assuming the circulation at the beginning is zero, 
\begin{equation} 
\label{eq:Kelvin_theorem}
    \sum_{j=1}^{N} \Gamma_{j} = 0,
\end{equation}
equation (\ref{eq:no_penetration}) and equation (\ref{eq:Kelvin_theorem}) provide $n+1$ linear equations to obtain the strength of all attached vortices $\Gamma_{j}, j =1,...n$, as well as the newly-shed free vortex element, $\Gamma_{N}$ at each time step.

The time evolution of the strength of the bound vortices changes the pressure jump across the fin $[p]$. The unsteady form of Bernoulli's equation \cite{eldredge2019mathematical} shows this relationship as \cite{alben2009simulating, jones2003separated}: 
\begin{equation}
\label{eq:bern_eqn}
\frac{\partial}{\partial t}\gamma(s,t)  + \frac{\partial}{\partial s} (v_{\text{rel}} \gamma(s,t) )=  \frac{\partial}{\partial s}[p](s,t), \quad 0 < s < 1
\end{equation}
where  $\gamma$ is the continuous vortex sheet strength on the fin and $ v_{rel}$ denotes the tangential velocity of the fluid relative to the fin, and $[p]$ is the pressure jump across the fin. Assuming pressure continuity at the trailing edge and integrating equation (\ref{eq:bern_eqn}) from the trailing edge to an arbitrary point $s$ along the fin, we obtain 
\begin{equation}
    \label{eq:bern_eqn1}
    \int_{1}^{s}\dot\gamma ds + v_{\text{rel}}\gamma - v_{\text{rel}} \gamma|_{s=1}  = [p](s).
\end{equation}
Using Kelvin's circulation theorem, we have $\int_{1}^{s}\dot{\gamma}\,ds
= \int_{0}^{s}\dot{\gamma}\,ds$,
and, at each time step $\left.v_{\mathrm{rel}}\gamma\right|_{s=1}= c$, where \(c\) is a constant. Assuming that both the vortex strength and the pressure jump are constant over each panel, Eq.~(\ref{eq:bern_eqn1}) can be discretized as
\begin{equation}
    \label{eq:bern_eqn2}
    [p]_k
    =
    \sum_{j=1}^{k}\Gamma_j
    +
    \frac{v_{\mathrm{rel},k}\Gamma_k}{l_k}
    + c,
\end{equation}
where \([p]_k\), \(v_{\mathrm{rel},k}\), and $l_k$ denote the average pressure jump and the average relative fluid velocity over the \(k^{\mathrm{th}}\) panel, and the length of the \(k^{\mathrm{th}}\) panel, respectively. After determining \(\Gamma_k\) from the previous step and computing the fluid velocity using the Biot-Savart law, the pressure jump over each panel can be obtained from Eq.~(\ref{eq:bern_eqn2}). The constant \(c\) is determined by enforcing pressure continuity at the trailing edge. Specifically, \(c\) is chosen such that the extrapolated pressure distribution function satisfies the unsteady form of the Kutta condition $[p]\big|_{s=1}=0$.

\subsection{Measures of performance}
\label{sec:performance}

\paragraph{Power expenditure} The instantaneous input power expended by a heaving and pitching flat plate during a particular `swimming strategy' is given by
\begin{equation} 
\label{eq:pin}
P_{\text{in}} = F_y \dot{h} + M_z \dot{\theta},
\end{equation} 
where $F_y$ is the lateral force integrated along the length of the flat plate, and $M_y$ is the pitching moment at the leading edge. For the tethered fin, we quantify the input power using the coefficient of power, $$C_P = \frac{P_{\text{in}}}{\frac{1}{2} U_0^3},$$ where $U_0$ is the free-stream speed. For the self-propelled swimmer, the power coefficient is no longer appropriate, and we directly use the time-averaged input power $\overline{P}_{\text{in}}$ to quantify the power expenditure. From now on we use the overbar symbol to indicate the time average of the corresponding quantity over one cycle. Note that the input power (\ref{eq:pin}) is itself a non-dimensional quantity since we have normalized lengths, times and masses by $L$, $2\pi/\omega$ and $\rho_f L^2$ respectively.

\paragraph{Thrust} For the tethered fin, we use the thrust coefficient $$C_T = \frac{T}{\frac{1}{2} U_0^2}$$ to quantify how successful a particular swimming strategy is, where $T$ is the upstream force (i.e., thrust) exerted by the fluid on the flat plate. For the self-propelled fin, the thrust coefficient is no longer appropriate because the speed of the swimmer is not an independent variable, and we instead use the steady forward speed of the fin-body system, $\overline{v}$ to quantify the thrust. When the swimmer achieves a steady forward speed, the thrust exerted by its caudal fin will equal the drag contributed by the virtual upstream body, i.e., $T = D = \mu v^2$. Therefore, the steady forward speed of a caudal fin-driven swimmer $\overline{v}$ indicates how much thrust is produced by a particular swimming strategy. Note that the speed is measured in units of `plate length per unit pitching/heaving period' and is therefore a non-dimensional quantity. Both $C_T$ and $\overline{v}$ may be negative for particularly unsuccessful swimming strategies.

\paragraph{Efficiency} 
To quantify the efficiency of both models of fish swimming, we use the Froude efficiency. For a tethered fin experiencing a free-stream speed $U_0$, the Froude efficiency can be understood as the ratio between the thrust coefficient and the power coefficient, i.e.,
\begin{equation}
    \label{eq:tethered_eta}
    \eta = \frac{C_T}{C_P} = \frac{T/\frac{1}{2}U_0^2}{P_{\text{in}}/\frac{1}{2}U_0^3} = \frac{T U_0}{P_{\text{in}}}.
\end{equation}
For a self-propelled swimmer in still fluid, the Froude efficiency depends on the average forward speed $\overline{v}$, since there is no free-stream speed. When the swimmer achieves a steady forward speed, the thrust equals $\mu v^2$, and the Froude efficiency for the self-propelled swimmer becomes
\begin{equation}
    \label{eq:free_eta}
    \eta = \frac{T v}{P_{\text{in}}} = \frac{\mu v^3}{P_{\text{in}}}.
\end{equation}

\subsection{Validation}
\label{sec:validation}

We validated the tethered model with the experimental results presented in \cite{VanBuren2019Scaling}. In the latter study, the authors examined the effect of changing $\phi$, the phase difference between simultaneous heave and pitch motion, for a tethered fin (chord length $c= 80$ mm, span $s = 279$ mm) with constant freestream velocity $U_0 = 0.1$ m/s and frequency $f = 0.2, 0.3, ..., 0.8$ Hz. We compared the performance of our tethered model with figures 6 and 7 of \cite{VanBuren2019Scaling} using the same reduced frequency $k \equiv {f c}/{U_{0}}$. 

\begin{figure}[h]
    \centering
    \includegraphics[width=0.6\textwidth]{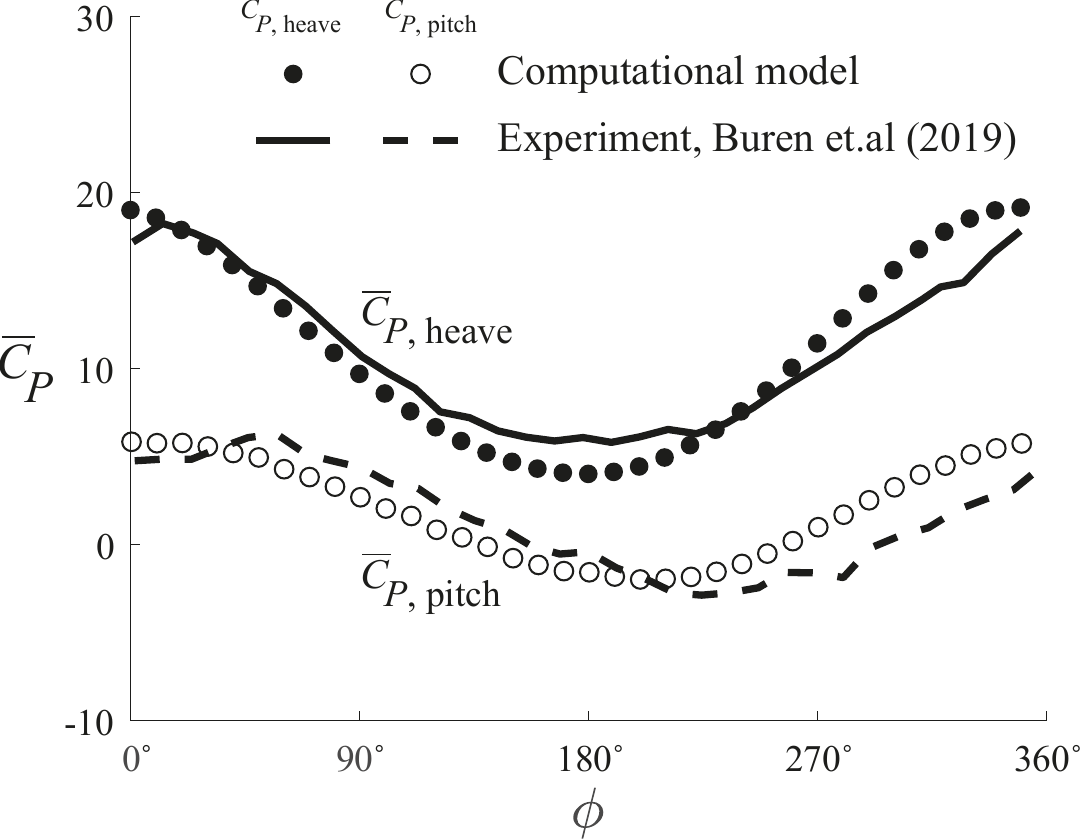}
    \caption{Comparison  of  coefficients of power $\overline{C}_{P,\text{pitch}}$ and $\overline{C}_{P,\text{heave}}$ between computational model and experimental results}
    \label{fig:validation}
\end{figure}

In figure \ref{fig:validation}, we show how the contributions of heaving and pitching motion to the coefficient of power, i.e., $$C_{p,\text{pitch}} = \frac{M_z \dot{\theta}}{\frac{1}{2} U_0^3}, \quad C_{p,\text{heave}} = \frac{F_y \dot{h}}{\frac{1}{2} U_0^3}$$ vary with the phase offset between pitch and heave $\phi$. The results from our computational model show close agreement with Van Buren et.~al's experimental measurements with a heaving and pitching airfoil in a water tunnel. We use the same computational model and thin airfoil theory for both the tethered and untethered models; the only difference between them is that in the self-propelled model, we assume that the fin is pushing a virtual body that does not affect the fluid dynamics.

\section{ Results and Discussion}

Numerical simulations were carried out for both a tethered and a self-propelled swimmer at 4 different pitch amplitudes, 3 different heave amplitudes, and 36 different phase offset values totaling 432 unique prescribed motions. In addition, for the tethered fin, six different uniform background flow speeds $U_0$ were imposed, whereas the self-propelled swimmer was simulated at three different values of the drag coefficient $\mu$. A list of the parameter values used in these simulations is given in table \ref{table:sim_matrix}. All quantities were calculated for 500 cycles of the fin's motion, and averaged over the last 100 cycles.

\begin{table}[h]
\caption{\label{table:sim_matrix}Experimental parameters for simulation study }
\begin{center}
\begin{tabular}{p{3.5cm}| p{4.5 cm}}
\hline
Pitch & $\theta_{o}$ = 5$^{\circ}$, 10$^{\circ}$, 15$^{\circ}$, 20$^{\circ}$  \\
\hline
Heave & ${h_{o}}$ = 0.125, 0.25, 0.375 \\
\hline
Phase & $\phi$  = 0$^{\circ}$, 10$^{\circ}$,  20$^{\circ}$, ... , 350$^{\circ}$   \\
\hline
Frequency & $f$  = 1   \\
\hline
Speed & $U_{0} = 0.25, 0.5, 1, 2, 5, 10 $     \\
\hline
Drag coefficient & $\mu = 0.025, 0.1, 0.2 $     \\
\hline
\end{tabular}
\end{center}
\end{table}

\subsection{Tethered fin } 
\label{subsec: Tethered fin}
%%% text added Emad
\subsubsection{Effect of phase offset $\phi$ on performance}

\begin{figure*}
\centering
{\includegraphics[width = 1\textwidth ]{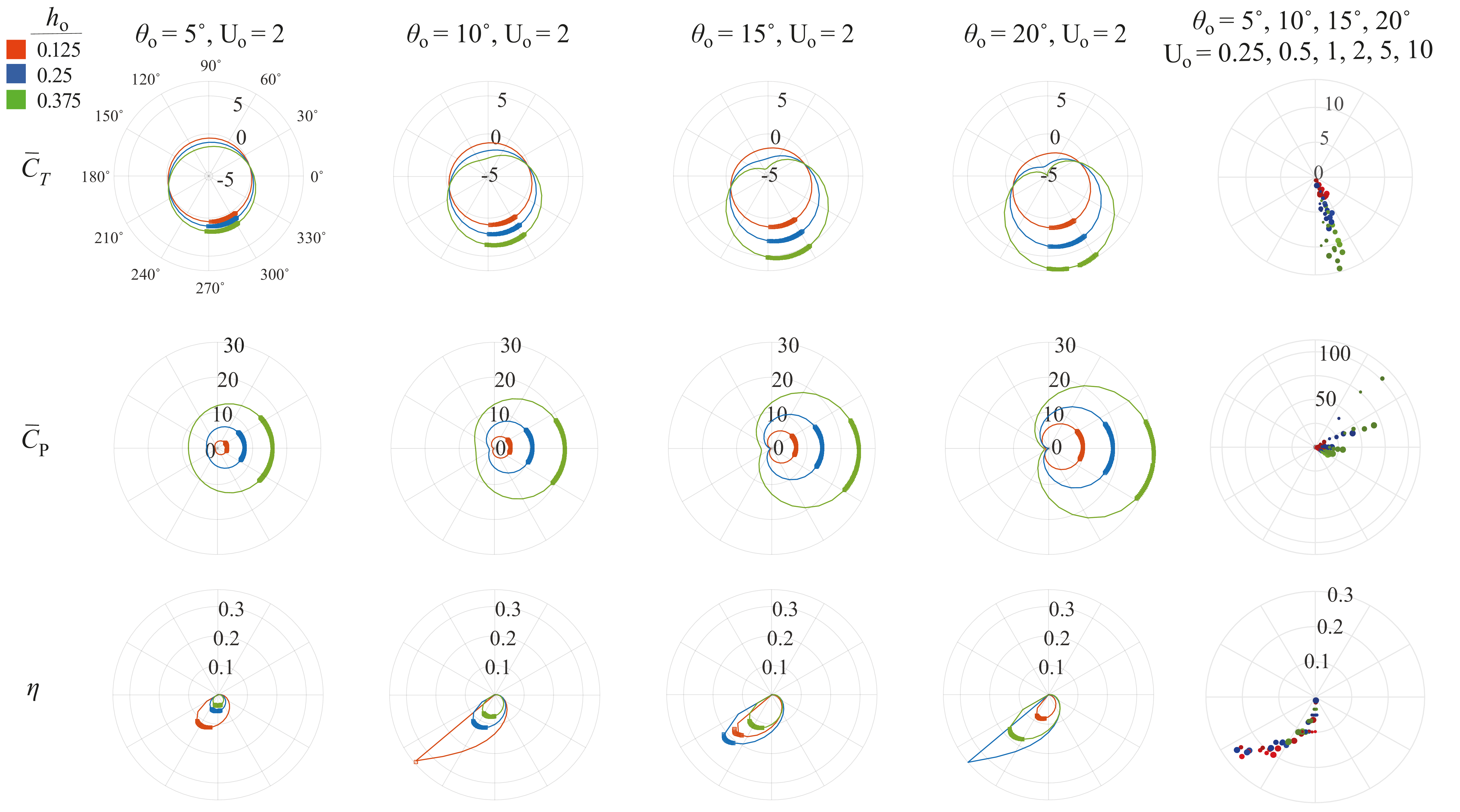}}
\caption{Variation of coefficient of thrust,  $\overline{C}_{T}$, coefficient of power $\overline{C}_{P}$, and efficiency $\eta$ (all shown in radial direction) for various heave, pitch, and phase offset (shown in the azimuthal direction). {When $\overline{C}_{T}<0$, the efficiency is reported as zero.} The last column summarizes the results for $U_0 = 0.25, 0.5, 1, 2, 5,$ and $10$. The size of the dots indicates the magnitude of the pitch, while the intensity of the colors indicates the magnitude of the background flow speed, with darker colors representing higher speeds. Each point in the last column represents the average of top $5\%$ of $\overline{C}_T$, $\overline{C}_P$, and $\eta$ for the given kinematic parameters}
\label{fig:tethered_CTCPeta_phi}
\end{figure*}

We found that the thrust coefficient can be either positive --- indicating a net forward (upstream) hydrodynamic force on the fin --- or negative, i.e. a net reverse (downstream) hydrodynamic force on the fin, depending on the value of $\phi$ (figure \ref{fig:tethered_CTCPeta_phi}). Across different heave and pitch amplitudes, negative thrust is obtained for $50\deg < \phi < 200\deg$, and the largest forward thrust is achieved for values of $270\deg < \phi < 300\deg$. This result agrees well with the experiments on a tethered NACA 0012 airfoil at high Reynolds number conducted by \cite{anderson1998oscillating}, who found that a phase offset of $\phi \approx 285 \deg$ optimizes the thrust. At this value of $\phi$, the thrust coefficient was not found to be very sensitive to changes in $\phi$, and the variation of the coefficient of thrust for $270\deg < \phi < 300\deg$ was found to be less than 5\%. We also found that the thrust increases with both pitch and heave amplitude; the maximum coefficient of thrust occurred at the highest pitch amplitude ($\theta_{o} = 20\deg$) and the highest heave amplitude ($h_{o} = 0.375$) we investigated.

The coefficient of input power, $\overline{C}_P$ (figure \ref{fig:tethered_CTCPeta_phi}, second row), is always positive and increases noticeably with an increase in the heave amplitude $h_o$ for any given pitch amplitude. The minimum $\overline{C}_P$ --- i.e., the least amount of power expended in moving the foil --- occurs at $\phi \approx 180\deg$, when the pitch and heave motions are exactly out of phase. This corresponds to the smallest trailing edge deflection amplitude for any given combination of $h_o$ and $\theta_o$, showing that power consumption is minimized when the body is aligned with the flow and has the smallest frontal area. On the other hand, $\overline{C}_P$ is maximized --- i.e., the most power is expended --- at $\phi \approx 0\deg$, when the pitch and heave motions are exactly in phase, leading to the largest possible trailing edge deflection amplitude for any given combination of $h_o$ and $\theta_o$. It is worth noting that there is an order of magnitude difference between the $\overline{C}_P$ at $\phi = 0\deg$ and at $\phi = 180\deg$.

The efficiency $\eta$ of an oscillating tethered fin, given by equation (\ref{eq:tethered_eta}) and shown as a function of $\phi$, $h_o$ and $\theta_o$ in the third row of figure \ref{fig:tethered_CTCPeta_phi}, can be interpreted as a combination of the thrust coefficient $\overline{C}_T$ and the input power coefficient $\overline{C}_P$: in other words, a highly efficient motion is one that maximizes $\overline{C}_T$ while minimizing $\overline{C}_P$. Unlike thrust and power, we observed no monotonic trend in $\eta$ when the pitch and heave amplitudes were increased. For instance, for large  heave amplitudes, $h_{o} = 0.25$ and 0.375, the efficiency increases with an increase in pitch amplitude. In contrast, at a low heave amplitude, $h_{o} = 0.125$, efficiency first increases and then decreases with an increase in pitch amplitude. However, across almost all values of $h_o$ and $\theta_o$, $\eta$ is maximized when the phase offset between pitch and heave is $220\deg < \phi < 270\deg$.

Thus, we found that for a tethered fin, efficiency and thrust are maximized by different ranges of phase offset values; what is optimal with reference to $\overline{C}_T$ is not necessarily optimal for $\eta$, and vice versa. In fact, the efficiency is maximized at a value of $\phi$ for which the thrust coefficient (as well as the input power coefficient) is quite small. {The `spikes' in the efficiency for $\theta_o =10 \deg$ and $\theta_o = 20 \deg$ at $\phi = 220$ seen in fig.~\ref{fig:tethered_CTCPeta_phi} are an artifact of our decision to report $\eta = 0$ when $C_T<0$ therefore due to the extremely low values of $\overline{C}_T$ and $\overline{C}_P$ for this value of the phase offset $\phi$.}

\begin{figure}
\centering
{\includegraphics[width = 1\textwidth ]{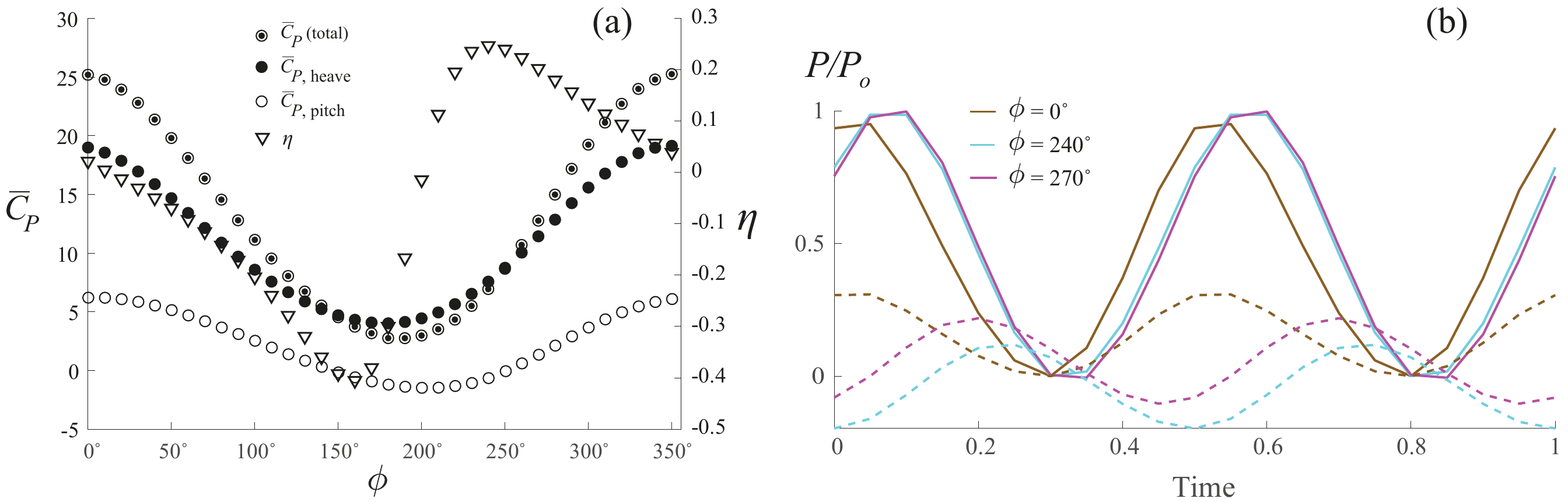}}
\caption{(a) Total coefficient of power, $\overline{C}_{P}$, coefficient of power due to pitch,  $\overline{C}_{P,\text{pitch}}$, and coefficient of power due to heave, $\overline{C}_{P,\text{heave}}$, together with efficiency $\eta$ as a function of phase offset between pitch and heave $\phi$. (b) Normalized heave power, $F_{y} \dot{h}$ (solid lines) and  pitch power, $M_{z}\dot{\theta}$ (dashed lines)  over one flapping cycle at $\phi$ = 0$^{\circ}$, 240$^{\circ}$, 270\deg$^{\circ}$. All values shown correspond to a pitch amplitude of $\theta_o = 15^{\circ}$ and heave amplitude of $h_o = 0.375$. A similar trend is observed for the other pitch and heave amplitudes.}
\label{fig:tethered_CPpitchvsheave}
\end{figure}

\subsubsection{Contribution of pitch and heave to the total power consumption:}
We found that for a tethered fin undergoing simultaneous periodic pitch and heave motion, heaving  always contributes a net positive power, i.e., the heave motion always \emph{consumes} power and expends energy \emph{into} the fluid (figure \ref{fig:tethered_CPpitchvsheave}). On the other hand, the pitching motion sometimes contributes a net negative power, i.e., pitching motion sometimes \emph{gains} power and therefore draws energy \emph{from} the fluid. Figure \ref{fig:tethered_CPpitchvsheave}(a) shows the total coefficient of input power, $\overline{C}_{P_{\text{, total}}}$, broken down into $\overline{C}_{P_{\text{, heave}}}$ and $\overline{C}_{P_{\text{, pitch}}}$, as a function of the phase offset between pitch and heave $\phi$ for a pitch amplitude of $\theta_0 = 15^{\circ}$ and a heave amplitude of $h_0 = 0.375$. It can be seen that over the range $150<\phi<250$, $\overline{C}_{P_{\text{, pitch}}}$ is negative, whereas $\overline{C}_{P_{\text{, heave}}}$ is always positive. This is broadly consistent with the experimental results of \cite{VanBuren2019Scaling}, who found that $\overline{C}_{P_{\text{, pitch}}}$ has a global minimum at $\phi \approx 270\deg$. The right-hand axis of figure~\ref{fig:tethered_CPpitchvsheave}(a) shows the efficiency $\eta$ as a function of $\phi$; this value can be negative or positive depending on the sign of the thrust coefficient $C_T$ (not shown).

For a simultaneously heaving and pitching tethered fin, the coefficient of total power, $\overline{C}_{P}$, is therefore minimized by a combination of low (positive) power consumption from heaving and negative power `consumption' from pitching during peak efficiency. To illustrate this phenomenon we plotted the power consumption due to pitch and heave over a single cycle of the fin's motion for three values of phase offset $\phi$: 0\deg, $240\deg$ and $270\deg$  in figure \ref{fig:tethered_CPpitchvsheave}(b). The contributions due to pitch and heave are both normalized by the maximum instantaneous power consumption over one oscillation cycle ($P_o = \text{max} \{ M_z \dot{\theta} + F_y \dot{h}\}$).
The result shows that the power drawn by heave dominates, and that the power drawn due to pitching motion is significantly lower compared to total power consumption. This, too, is consistent with the experimental results of \cite{VanBuren2019Scaling}, who also found that the power consumption for a simultaneously pitching and heaving foil is largely due to $\overline{C}_{P\text{, heave}}$, and that the power consumption from pitch can even be neglected for scaling purposes.

Figure \ref{fig:tethered_CPpitchvsheave}(b) shows that at $\phi = 0\deg$, which is a particularly inefficient prescribed motion for the tethered fin, the power consumption  from the pitching motion (dashed brown) remains positive for the entire cycle of motion. Since $\overline{C}_{P_{\text{, heave}}}$ is always positive, their combined effect is to develop a high total coefficient of power, as indicated in the peak in the power coefficient at $\phi = 0 \deg $ in figure~\ref{fig:tethered_CPpitchvsheave}(a). %Although there are points in the fin's motion when the combined pitch and heave did not consume any power (instantaneously), over the entire cycle both $\overline{C}_P_{\text{, heave}}$ and $\overline{C}_P_{\text{, pitch}}$ add up to positive values. 
On the other hand, at $\phi = 240\deg$, which corresponds to the highest-efficiency case, we found that the relative power consumption from pitch (dashed blue in fig.~\ref{fig:tethered_CPpitchvsheave}(b)) becomes significantly lower than when $\phi=0\deg$  and, crucially, negative for a significant portion of the cycle. Thus, our results indicate that efficient oscillatory motion of a tethered fin is due to low or even negative power consumption from the pitching component of its motion.

\subsection{Self-propelled model} 
\label{subsec:freelyswimmingfin}
Triantafyllou and others have found using experiments on a tethered swimmer that a phase lag of $\phi = 270\deg$ maximizes the efficiency of a tethered swimmer \cite{anderson1998oscillating,barrett1999drag,read2003forces}, similar to what we have found in section 3.1. Perhaps due to the influence of Triantafyllou's work, many numerical studies of self-propelled swimmers since then have been conducted at a fixed value of $\phi = 270\deg$ \cite{hover2004effect,singh2008hydrodynamics,paniccia2021performance,akoz2021intermittent,paniccia2022locomotion}. A comprehensive analysis of the effect of $\phi$ on the performance of a fish-like body has, to our knowledge, only been conducted for tethered swimmers, most recently in \cite{fernandez-feria_note_2017,VanBuren2019Scaling,matthews_role_2022}. In this section, we report the results of a parametric study of the propulsive performance of a caudal-fin driven fish that is free to move in the streamwise direction, allowing us to explicitly compare the performance of a self-propelled swimmer with a tethered one.

As outlined in section \ref{sec:performance}, the same performance measures cannot necessarily be used for the two types of swimmers, because a self-propelled swimmer in still fluid achieves a steady speed $\overline{v}$ that is an outcome of a particular swimming strategy rather than an independently controlled parameter. Therefore, the thrust and power coefficients --- for which the speed appears in the denominator --- cannot be used to quantify a self-propelled swimmer's performance. Instead, we have used the time-averaged forward speed $\overline{v}$ to quantify how successful a self-propelled swimmer's heaving/pitching strategy is, and we have used the time-averaged input power (\ref{eq:pin}) in place of the power coefficient to quantify the power expenditure.

Even though $P_{\text{in}}$ and $C_P$ cannot be numerically compared against each other, we believe that this offers a consistent way to qualitatively compare the power expenditure in the tethered versus the self-propelled configurations. Similarly, while the thrust coefficient and steady forward speed cannot be quantitatively compared against each other numerically, $C_T$ and $\overline{v}$ can nevertheless be used to answer questions such as: what swimming strategy maximizes the `performance' of a tethered fin, and does the same swimming strategy also maximize the `performance' of a self-propelled swimmer?

The Froude efficiency can straightforwardly be used for the self-propelled swimmer, allowing direct numerical comparison with the Froude efficiency of a tethered fin from section \ref{subsec: Tethered fin}. In this section, we use the efficiency from eq. (\ref{eq:free_eta}), which can be considered to be the ratio between the useful power output of the caudal fin and the power expended by the pitch/heave motion.

Using the parameters shown in table \ref{table:sim_matrix}, a self-propelled swimmer was modeled for the same set of kinematic strategies as the tethered fin. In addition, three values of the drag coefficient $\mu = \{ 0.025,0.1,0.2 \}$ were also considered. For each set of parameters, the model was run for $500$ cycles of the prescribed kinematics until steady-state motion was reached. All performance parameters were then measured using time-averages over the last 100 cycles of fin motion.

\begin{figure*}[ht]
\centering
{\includegraphics[width = 0.9\textwidth] {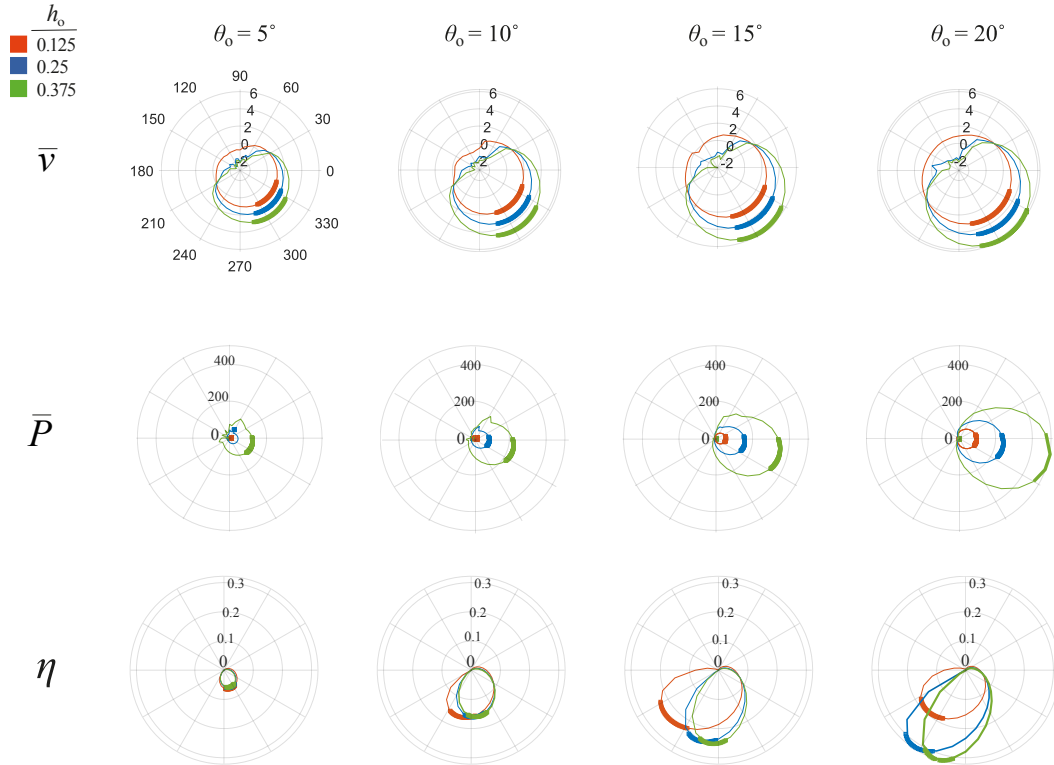}}
\caption{Variation of  steady-state speed ($\overline{v}$), power consumed by the fin ($\overline{P}$), and efficiency ($\eta$) with phase offset for various pitch and heave amplitudes. Highlighted region in each plot represents values within the top 5\%. In these simulations $\mu = 0.1$.} 
\label{fig:free_CTCPeta_phi}
\end{figure*}

\subsubsection{Effect of phase offset $\phi$ on performance:}
\label{subsubsec:freeperformance}
%\subsubsection{{Effect of phase offset on average speed} \label{subsubsec:speed}}

As one would expect, the magnitude of the self-propelled speed increases monotonically with both pitch and heave amplitudes (figure \ref{fig:free_CTCPeta_phi}, first row), and reaches values of nearly 6 chord lengths per cycle of oscillation when $\{\theta_{o}, { h_{o}}\}$ = \{20$^{\circ} , 0.375$\}. The phase offset $\phi$ has a significant effect on the swimming speed, with the highest forward speeds developed in the range $290\deg < \phi < 330\deg$. There is also a large range of $\phi$ which is unfavorable for free swimming: in the range $50< \phi < 200$, the fin swims backward or oscillates in its original place with a very small net displacement. Recall that the thrust coefficient $\overline{C}_T$ for a tethered fin was found in this work to peak when $\phi$ was between 270\deg and 300\deg, in agreement with published experimental and numerical studies by \cite{anderson1998oscillating} and \cite{read2003forces}. Comparing the tethered and untethered cases, we find that the phase offset corresponding to maximum $\overline{C}_T$ for a tethered fin is not the same as the phase offset corresponding to maximum $\overline{v}$ for a self-propelled. $C_T$ for a tethered fin peaks at $\phi = 285\deg$, whereas $\overline{v}$ for a self-propelled peaks at $\phi = 310\deg$. This shows that while there are broad similarities between tethered swimmers and self-propelled swimmers undergoing the same kinematics, it cannot be assumed that the same set of kinematic parameters that maximize the thrust in the tethered case will also maximize the steady-state speed in the self-propelled case in general.

The total work done by the fin also increases with both pitch and heave amplitudes. Overall, there is a marked resemblance between the polar plots of $\overline{P}$ in the self-propelled case (figure ~\ref{fig:free_CTCPeta_phi}, second row) and $C_P$ in the tethered case (figure~\ref{fig:tethered_CTCPeta_phi}, second row), which is to be expected since the two quantities are both measures of the power consumption due to the same kinematic strategy. The range of $\phi$ over which a self-propelled swimmer does the largest amount of work per cycle, $330\deg < \phi < 360 \deg$, is different from the range of $\phi$ over which a tethered fin has the greatest coefficient of input power, $-30\deg < \phi < 30 \deg$. However, almost no work is done by the fin at values of $\phi$ between $90\deg$ and $270\deg$, which is consistent with what was observed for the power coefficient of a tethered fin. 

The third row of figure \ref{fig:free_CTCPeta_phi} shows that the efficiency, $\eta$, is maximized in the range $220\deg < \phi < 290\deg$ for a self-propelled swimmer. For small pitch amplitudes, $\eta$ is maximized by $\phi \approx 290\deg$. As the pitch amplitude is increased, we find that the  $\phi$ corresponding to peak efficiency decreases; a similar trend was observed in the Froude efficiency of a tethered fin, as well. The highest efficiency of 0.32 is achieved by the largest values that we considered, $\{\theta_{o}, { h_{o}}\}$ = \{20$^{\circ} , 0.375$\}, suggesting that combined large-amplitude heave and pitch motion leads to particularly efficient self-propulsion. Interestingly, for small pitch amplitudes $\theta_o$, $\eta$ was not found to be particularly sensitive to the heave amplitude $h_o$; for example, at $\theta_{o} = 10\deg$, a maximum $\eta$ of approx. 0.25 was reached regardless of whether the heave amplitude was $0.125$ or $0.375$. This was not the case for a tethered fin, whose maximum efficiency depended significantly on $h_o
$, as shown in the third row of figure ~\ref{fig:tethered_CTCPeta_phi}.

\subsubsection{Pitch and Heave Work:}

Similar to the tethered fin considered earlier, it is instructive to break down the average work done per oscillation cycle by the self-propelled swimmer into pitch and heave work. Just as in the tethered case, we find that pitching does negative work for some values of $\phi$, i.e., it draws energy from the surrounding fluid, whereas heaving always does positive work regardless of the phase offset between pitch and heave (figure \ref{fig:free_CPpitchvsheave}(a)); the total amount of work then remains positive for all values of $\phi$. In the self-propelled case, we find that negative work from pitch occurs over the range $180 \deg < \phi < 250 \deg$, which is similar to, but not the same as, the corresponding range of $\phi$ for which a tethered fin does negative work ($150 \deg < \phi < 250 \deg$).

\begin{figure}
\centering
{\includegraphics[width = 1\textwidth] {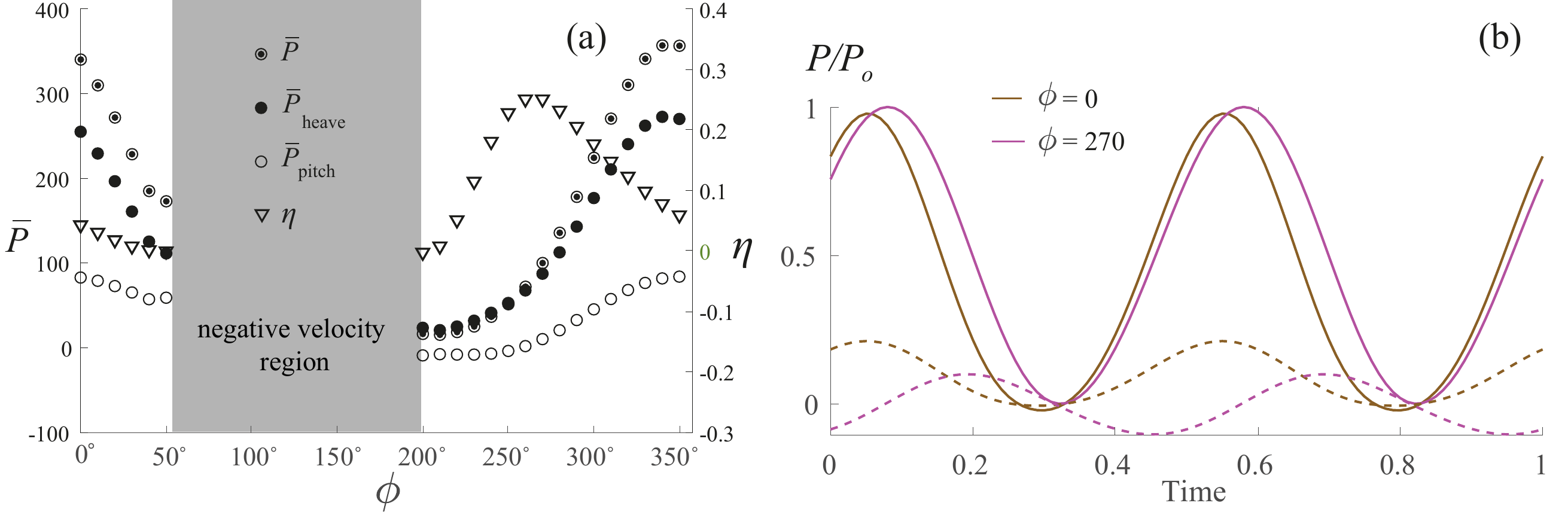}}
\caption{(a) Variation of total power consumption $\overline{P}$, power consumption due to the heave motion $\overline{P}_{heave}$, and pitch motion $\overline{P}_{pitch}$  and efficiency, $\eta$ with phase offset $\phi$; (b) Normalized pitch (dashed) and heave power (solid) for self-propelled swimmer at $\phi$ = 0$^{\circ}$, 270\deg$^{\circ}$ over one oscillation period. All values shown correspond to a pitch amplitude of $\theta_{o} = 10^{\circ}$ and heave amplitude of $ h_{o} = 0.375$; the drag coefficient is $\omega = 0.1$.} 
\label{fig:free_CPpitchvsheave}
\end{figure}

To further explore the influence of phase offset on peak efficiency, we studied the normalized pitch and heave power as a function of time for one oscillation period at two different phase offsets $\phi$ = $0\deg$ and $\phi =270\deg$. We found that at $\phi = 0\deg$ --- which is the phase offset corresponding to maximum work done by the fin --- the pitching motion does positive work for nearly the entire oscillation cycle. But at $\phi = 270\deg$ --- which corresponds to maximum efficiency, pitching does negative work for most of the oscillation cycle (figure~\ref{fig:free_CPpitchvsheave}(b)).  This implies that the surrounding fluid aids the work associated with pitching moment for $\phi = 270\deg$ .  Meanwhile, the work done by heaving motion does not change much between these two values of $\phi$, which shows that the improvement in efficiency at optimal values of $\phi$ is brought about through a change in the work done by pitching, not heaving.

\subsubsection{Effect of the body shape:}
 \label{subsubsec:mu}

\begin{figure*}[]
 \centering
   {\includegraphics[width = 0.9\textwidth] {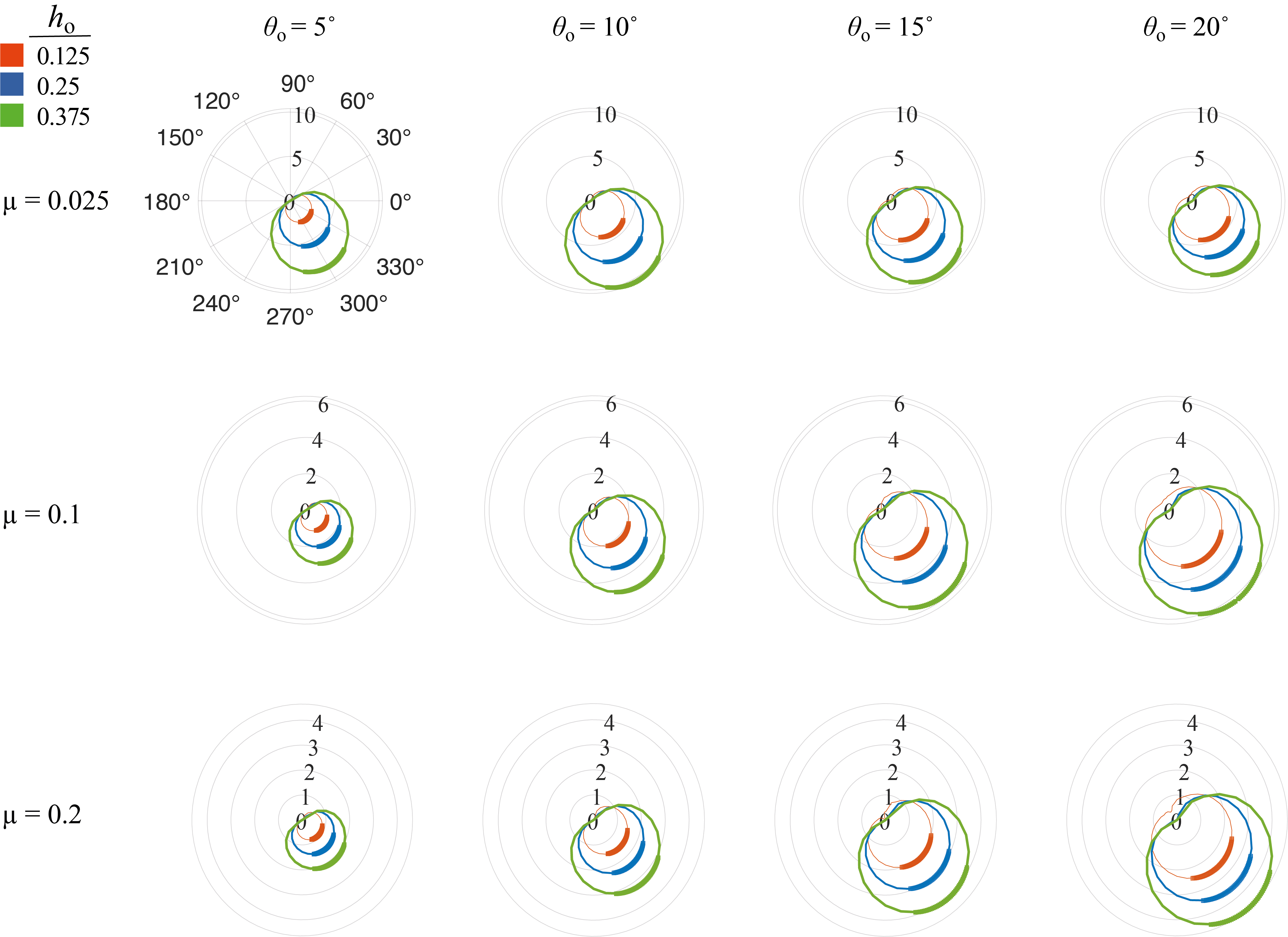}}
   \caption{Variation of  average speed,  $\overline{v}$    for various coefficient of drag, pitch angle, and heave amplitudes. The self-propelled swimmer is simulated for $\mu = 0.025, 0.1, 0.2$, $\theta_{o} = $5$^{\circ}$,10$^{\circ}$,15$^{\circ}$,20$^{\circ}$, and ${h_{o}} = 0.125, 0.25, 0.375$. Highlighted region in each plot represent $\overline{v} \geq 0.95 \ v_{max}$ respectively.} 
   \label{free:speed_vs_mu}
\end{figure*}

The main difference between the tethered and self-propelled models is that in the tethered case, the speed of the fluid is fixed, whereas in the self-propelled case, the speed of the swimmer is determined using Newton's second law applied to the fin and a `virtual body' attached to it. The combined thrust and drag of this fin-body system creates an unbalanced force on the self-propelled swimmer, causing it to accelerate until it reaches a steady speed when the thrust equals the drag. In this calculation, the drag coefficient $\mu$ is a key parameter, which depends on the shape of the upstream body and, experimentally, has been measured and reported in the literature for various species of fish. In this section, we briefly examine the effect of changing $\mu$ on the conclusions presented in the preceding sections. To do so, we use three different values of $\mu = \{0.025, 0.1, 0.2\}$ as shown in table \ref{table:sim_matrix} and analyze the performance of the self-propelled swimmer for the full set of kinematic parameters for each of these values of the drag coefficient. The questions we are interested in are: how does {a tail fin-driven swimmer's} performance change with {the upstream} body shape for a given {set of} kinematic parameters of the fin? How sensitive is the optimum $\phi$ to the {upstream }body shape?

The results indicate that the body shape does not appear to affect the {value of $\phi$ that optimizes the average speed $\overline{v}$} (Figure \ref{free:speed_vs_mu}).  The range of $\phi$ values {that leads to higher speeds is} highlighted, and {seems to be insensitive to variations in $\mu$}. This {suggests} that fish with different body shapes and drag coefficients could maximize their swimming speed using the same kinematics, i.e., using the same prescribed motion of the caudal fin. It will be instructive to investigate {this} result on a freely-swimming robot {in} a future {study}.

\section{Conclusion}
\label{sec:conclusion}
Many studies of fish swimming use a tethered configuration in which the swimmer is constrained from moving in the streamwise direction \cite{Seo2022-improved,Seo2026-scaling,White2021-tunabot,zhu2019tuna}. However, it is questionable whether the results of such experiments are applicable to self-propelled swimmers. In particular, it is an open question whether the swimming strategies that optimize the performance of a tethered swimmer are also the ones that optimizes the performance of a self-propelled swimmer. This question is made more complicated by the fact that the same definition of `performance' cannot be used for a tethered swimmer in a free stream and a self-propelled swimmer in quiescent fluid 

In this work, we have systematically explored the performance of a fin in both tethered (section 3.1) and self-propelled (section 3.2) conditions using the same kinematic strategy: a sinusoidally-varying pitch and heave motion at the leading edge, with a variable phase offset $\phi$ between the pitch and the heave. In the tethered case, the amount of thrust developed by the fin while it remains fixed in the streamwise direction is a useful measure of the performance of the swimmer. In the self-propelled case, the average thrust is assumed to be counteracted by an average drag force, and the drag is assumed proportional to $v^2$. Thus, for a given upstream body (i.e., a given drag coefficient $\mu$), the average velocity of the body indicates the thrust generated by the fin. Therefore, we have used the average propulsive speed $\overline{v}$ to quantify the performance of a self-propelled swimmer.

While there are broad similarities between the swimming performance of a tethered caudal fin and the swimming performance of the same caudal fin when attached to a drag-producing upstream body, this study shows that there are noticeable differences between the two. The swimming strategies that maximize thrust on a tethered fin at a fixed flow speed are not necessarily the same as the swimming strategies that maximize the speed of a self-propelled swimmer with a given body shape, i.e., drag coefficient. The power consumed by a tethered fin when executing a particular swimming strategy is, likewise, not necessarily the same as the power consumed by the same fin when it propels a drag-producing body through still fluid. Thus, the Froude efficiency of these two cases will also, in general, be different

In particular, we have investigated the effect of varying the phase offset $\phi$ between simultaneous pitch and heave at several different heave and pitch amplitudes. The question of what is the optimal $\phi$ for a fish simultaneously pitching and heaving its caudal fin has been extensively studied in the literature for both tethered and self-propelled swimmers, although most studies tend to focus on the former. Our numerical results indicate that there is an approximately 30-degree difference between the optimal $\phi$ for thrust generation on a tethered fin compared to the optimal $\phi$ for the speed of a self-propelled swimmer. Therefore, the results of experiments on tethered swimmers may not always be applicable to self-propelled swimmers. 

The results of this study also suggest that a fish that generates thrust with its caudal fin optimizes its cruising speed $\overline{v}$ at the same value of $\phi$ regardless of the drag coefficient $\mu$ assumed for the upstream body, at least within the range of $\mu$ considered here ($0.025, 0.1, 0.2$). This suggests that a large variety of carangiform swimmers with different body shapes (having correspondingly different drag coefficients $\mu$) may require similar swimming strategies if their goal is to maximize their cruising speed. However, further computational and experimental studies are needed to fully consider the effect of the body shape on swimming performance. In addition, it remains to be seen how our conclusions will be affected by three-dimensional effects and the flexibility of the caudal fin.

\section*{Bibliography}
\bibliographystyle{unsrt}
\bibliography{swimming}
\end{document}